\documentclass[a4paper,11pt]{article}
\usepackage{pos}
\usepackage{amsmath}
\def\be{\begin{equation}}
\def\ee{\end{equation}}
\newcommand{\bea}{\begin{eqnarray*}}
\newcommand{\eea}{\end{eqnarray*}}
\newcommand{\ii}{{\rm i}}
\newcommand{\dd}{{\rm d}}

\title{Poisson gauge theory: a review}

\author*[a,b]{Maxim Kurkov}

\affiliation[a]{\textit{\footnotesize Dipartimento di Fisica ``E. Pancini'', Universit\`a di Napoli Federico II,\\ Complesso Universitario di Monte S. Angelo Edificio 6, via Cintia, 80126 Napoli, Italy.}}

\affiliation[b]{\textit{\footnotesize INFN-Sezione di Napoli, \\ Complesso Universitario di Monte S. Angelo Edificio 6, via Cintia, 80126 Napoli, Italy.}}

\emailAdd{max.kurkov@gmail.com}

\abstract{
In this  paper we overview the Poisson gauge theory focusing on the most recent developments. We discuss the general construction and its symplectic-geometric interpretation. We consider explicit realisations of the formalism for all non-commutativities of the Lie algebraic type. We discuss Seiberg-Witten maps between  Poisson gauge field-theoretical models. 
}

\FullConference{%
  Corfu Summer Institute 2022 "School and Workshops on Elementary Particle Physics and Gravity",\\
  28 August - 1 October, 2022\\
  Corfu, Greece}

\begin{document}
\maketitle

\section{From non-commutative geometry to Poisson gauge theory.}
\noindent 
Various approaches to quantum gravity naturally yield non-commutative geometry, what motivates studies on non-commutative quantum field theory~\cite{Szabo:2001kg}. An important class  of field theories on non-commutative space-times is given by non-commutative \emph{gauge} theories, see~\cite{Hersent:2022gry} for the latest review. 

Let $\mathcal{M}$ be a flat manifold which represents the space-time. The local coordinates on $\mathcal{M}$ are denoted through $x^{a}$, $a=0,...,n-1$.
The non-commutative structure of the space-time is encoded in the  Kontsevich  star product\footnote{Of course, the star product is not the only way to treat the non-commutative geometry, for example, there is the spectral approach, for a review see~\cite{Devastato:2019grb}.  } of smooth functions on $\mathcal{M}$,
\be
 f(x) \star g(x) = f(x)\cdot g(x) + \frac{\ii}{2}\{f(x),g(x)\} + ... \,\, , \quad\quad \forall f,g\in \mathcal{C}^{\infty}\big(\mathcal{M}\big).
 \ee
In this formula the Poisson bracket is related to the Poisson bivector $\Theta^{ab}$,
 \begin{equation}
\{f(x),g(x)\}=   {\Theta^{ab}(x)}\,\,\partial_a f(x)\,\,\partial_b g(x)\, 
, 
\end{equation}
whilst the remaining terms, denoted through "...", contain second and higher derivatives of $f$ an~$g$.
 
 According to the novel approach to non-commutative gauge theories, proposed in~\cite{Kupriyanov:2020sgx}, non-commutative deformations of the $U(1)$ gauge theory must exhibit the following fundamental properties.
 \begin{itemize}
 \item{
 The infinitesimal gauge transformations should close the non-commutative algebra, 
 \be
 [ \delta_f,\delta_g ] A_{a}(x)
= \delta_{-\ii[f(x),g(x)]_\star}A_{a}(x)
\, . \label{ncgalg}
 \ee
 }
 \item{
 The commutative limit has to reproduce the usual abelian gauge transformations,
\be
\lim_{\Theta\to 0} \delta_f A_{a}(x) = \partial_{a} f(x) \,.  \label{colim}
\ee
}
\end{itemize}
Upon the semi-classical limit the star commutator, 
 \be
[f(x), g(x)]_\star=   f(x)\star g(x) - g(x)\star f(x)  = \ii \{f(x),g(x)\} + ...\,\,, \nonumber
\ee
tends to $\ii \{f(x),g(x)\}$, and the full non-commutative algebra~\eqref{ncgalg} reduces to the Poisson gauge algebra,
\textcolor{black}{
\be
[\delta_f,\delta_g]A_{a}(x)=\delta_{\{f(x),g(x)\}}A_{a}(x)\,.  \label{poga}
\ee}
The corresponding gauge theory is called the \emph{Poisson gauge theory}~\cite{Kupriyanov:2021aet}.
Physically this theory provides a semi-classical approximation of the non-commutative $U(1)$ gauge theory. 

The rest of this paper is organised as follows. In Sec.~2 we describe the general construction of  Poisson gauge models. Sec.~3 is devoted to its geometric interpretation. In Sec.~4 we present explicit formulae for non-commutativities of the Lie algebraic type. In Sec.~5 we consider the arbitrariness of the construction and its relation to Seiberg-Witten maps between Poisson gauge models. We round up with the summary and concluding remarks in Sec.~6.

\section{General construction. }  
\subsection*{a. Deformed gauge transformations.}
\noindent For the canonical non-commutativity, i.e. when
 $\partial_a\Theta^{bc} =0$, the required deformed gauge transformations can be easily constructed,
\begin{equation}
\delta_f A_a(x)=\partial_a f(x)+ \{A_a(x),f(x)\}\,. \label{infpgt}
\end{equation}
Though
for non-constant Poisson bivectors $\Theta^{ab}$ this simple formula is not compatible with the Poisson gauge algebra~\eqref{poga}, it can be generalised in the following way~\cite{Vlad,Kupriyanov:2020sgx},
\begin{equation}
\delta_f A_a(x) =\gamma^b_a(x,A(x))\,\partial_b f(x)  +\{A_a(x) ,f(x)\}\, . \label{infgtr}
\end{equation}
The matrix $\gamma(x,p)$, entering this relation, is a solution of the \emph{first master equation,}
\be
 \gamma_{a }^{ b } \partial^{ a }_p \gamma^{ d}_{ c} - \gamma^{ d}_{ a } \partial_p^{ a } \gamma^{ b}_{ c} 
+ \Theta^{b a} \partial_{ a } \gamma^{ d}_{ c} 
- \Theta^{ d a} \partial_{ a } \gamma^{ b }_{ c} 
- \gamma^{ a }_{ c}\partial_{a }\Theta^{ b  d} = 0,\quad \partial_p^c \equiv \frac{\partial}{\partial p_c }\,, \label{master1}
\ee
which tends to the identity matrix at the commutative limit,
\be
\qquad \lim_{\Theta\to 0}\gamma_{ b }^{ a } = \delta_b^{a} . \label{gcomlim}
\ee
One can check by direct calculation that these deformed gauge transformations, indeed, close the required non-commutative Poisson gauge algebra~\eqref{poga} and exhibit the correct commutative limit~\eqref{colim}.

\subsection*{b. Deformed field strength.}
\noindent According to the general strategy of~\cite{Kupriyanov:2020sgx}, the deformed field strength has to transform in a covariant way, 
\be
\delta_f {\cal F}_{ab}(x) = \{{\cal F}_{ab}(x),f(x)\}\,, \label{gcovcond}
\ee
and it has to reproduce  the commutative limit correctly,
\be
 \lim_{\Theta\to 0} {\cal F}_{ab}(x) = \partial_a A_b(x) -\partial_b A_a(x)   \,.
\ee
Such a field strength has been obtained in~\cite{Kupriyanov:2021aet} in the following form,
\begin{equation}
\mathcal{F}_{ab}(x) =\rho_a^{c}(x,A(x))\, \rho_b^{d}(x,A(x))\,\big(\gamma_c^l(x, A(x))\,\partial_lA_d-\gamma_d^l(x,A(x))\,\partial_lA_c+\{A_c \,, A_d\} \big)\, . \label{gcovfist} 
\end{equation}
The matrix $\rho(x,p)$, which enters this formula, has to obey the \emph{second master equation},
\be
 \gamma^c_d \partial_p^d\rho_a^b + \rho_a^d \partial_p^b\gamma_d^c + \Theta^{cd}\partial_d\rho^b_a = 0 \,, \label{master2}
 \ee
 and it has to recover the identity matrix at the commutative limit,
 \be
 \lim_{\Theta\to 0} \rho^{a}_b  = \delta_b^a\,.  \label{rhocomlim}
\ee

\subsection*{c. Gauge-covariant derivative.}
\noindent For any field $\psi(x)$, which transforms in a covariant way, 
\be
\delta_f\psi(x):=\{f(x),\psi(x)\}\,, 
\ee
the gauge-covariant derivative has to transform properly,
\be
\delta_f\left({\cal D}_a\psi(x)\right) = \{{\cal D}_a\psi(x),f(x)\}\, ,
\ee
and it has to exhibit a correct commutative limit,
\be
\lim_{ {\Theta}\to0}{\cal D}_a\psi(x) = \partial_a\psi(x)\,.
\ee
A suitable gauge-covariant derivative has been constructed in~\cite{Kupriyanov:2021aet} as follows,
\be
{\cal D}_a\psi(x) = \rho_a^i(x, A(x))\,\big(\gamma^l_i(x,A(x))\,\partial_l \psi(x) + \{A_a(x),\psi(x)\} \big) \, . \label{gcovdiv}
\ee
\subsection*{d. Gauge-covariant equations of motion.}
\noindent The deformed covariant derivative~\eqref{gcovdiv} and the deformed field strength~\eqref{gcovfist} allow to write down the “natural" equations of motion (e.o.m.),
 \be
{\cal D}_a {\cal F}^{ab} =0\, , \label{nateqm}
 \ee
which are manifestly gauge-covariant,
\be
\delta_f\left({\cal D}_a {\cal F}^{ab}\right) = \{{\cal D}_a {\cal F}^{ab},f\}, \label{gcovcondbis}
\ee
and which recover the first pair of Maxwell's equations in vacuum at the commutative limit. 
The gauge-covariance condition~\eqref{gcovcondbis} insures that the infinitesimal gauge transformations~\eqref{infgtr} map solutions of the e.o.m.~\eqref{nateqm} onto other solutions.

For the three-dimensional space-time, equipped with the $\mathfrak{su}(2)$ non-commutativity\footnote{Various aspects of the $\mathfrak{su}(2)$ non-commutativity have been studied in~\cite{Vitale:2012dz, Vitale:2014hca,Gracia-Bondia:2001ynb}.  }, these “natural" equations of motion are equivalent to the Euler-Lagrange equations, which come out from the gauge-invariant classical action,
\be
S[A] = \int_{\mathcal{M}} \dd x\,\left(-\frac{1}{4} \mathcal{F}_{ab}\mathcal{F}^{ab}\right),
\ee
see~\cite{Kupriyanov:2022ohu} for details. A more profound discussion on the gauge-covariant equations of motion can be found in~\cite{NewThoughts}.

\section{Symplectic geometric interpretation.}
\noindent The symplectic geometric interpretetion of the Poisson gauge formalism, designed in~\cite{Kupriyanov:2022ohu,Kupriyanov:2021cws}, is based on the two pillars:
the symplectic embedding, and the set of constraints in the outcoming symplectic space.
Extending the Poisson structure from $\mathcal{M}$ to $T^*\mathcal{M}$,
\begin{equation}
\{x^a,x^b\}=\Theta^{ab}(x)\,,\qquad\{x^a,p_b\}=\gamma^a_b(x,p)\,,\qquad \{p_a,p_b\}=0\,,
\end{equation}
we construct the symplectic embedding. One can easily see that the first master equation~\eqref{master1} for $\gamma$ is equivalent to the Jacobi identity on $T^*\mathcal{M}$.
By definition, the set of  constraints in this symplectic space reads,
\begin{equation}
\Phi_a(x,p):=p_a-A_a(x)\label{Phi}, \quad\quad a =0,...,n-1\,.
\end{equation}

The constituents of the Poisson gauge formalism exhibit natural expressions in terms of this symplectic geometric construction.
In particular, 
the deformed gauge transformation~\eqref{infpgt}  can be represented through a simple Poisson bracket on $T^*\mathcal{M}$,
\begin{equation}
\delta_f A_a(x) 
= \{f(x),\Phi_a(x,p)\}_{\Phi(x,p) =0} \,.
\end{equation}
The deformed field strength~\eqref{gcovfist} and the gauge-covariant derivative~\eqref{gcovdiv} can also be represented in a similar manner, 
\be
{\cal F}_{ab}(x) = \{\Phi^\prime_a(x,p),\Phi^\prime_b(x,p)\}_{\Phi^\prime(x,p)=0}\,, \qquad
{\cal D}_a\psi(x) = \{\psi(x),\Phi^\prime_a(x,p)\}_{\Phi^\prime(x,p)=0} \,,
\ee
where a new set of the constraints is defined as follows,
\be
\Phi^\prime_a(x,p):=\rho_a^b(x,p)\,\Phi_b(x,p) \,.
\ee
We remind that $\rho^a_b$ obeys the second master equation~\eqref{master2} and exhibits the commutative limit~\eqref{rhocomlim}.
Since $\rho$ is a non-degenerate matrix, the new and the old constraint surfaces coincide,
\be
\Phi^\prime_a(x,p)=0\,,\quad\Leftrightarrow\quad \Phi_a(x,p)=0\,. 
\ee

\section{Lie algebraic non-commutativities and universal solutions.}
\noindent Consider a class of Poisson bivectors, which are linear in coordinates,
\be
 \Theta^{ab} =  f^{ab}_{c}\, x^c \,.\nonumber
\ee 
The constants $f^{ab}_{c}$ satisfy the Jacobi identity, 
\be
f^{ec}_{a} f^{bd}_{c}  + f^{b c}_{a} f^{de}_{c} + f^{dc}_{a}f^{eb}_{c} = 0\, , \nonumber
\ee
therefore these constants can be seen as the structure constants of a Lie algebra. For this purpose we say that the corresponding 
non-commutativities are of the Lie algebraic type. 

In~\cite{Kupriyanov:2021cws} a special solution of the first master equation~\eqref{master1} has been presented in terms of a single matrix-valued function, which is valid for \emph{all} Poisson bivectors of the Lie algebraic type.
Introducing the matrix\footnote{According to our notations, for any matrix $B$ the upper index enumerates strings, whilst the lower one enumerates columns. In particular for a product of two matrices $B$ and $C$ we write
$
[B\,C]^{i}_j = B^i_k \,C^k_j.
$}, 
\be
[\hat{p}]^b_c= -\ii f^{ab}_{c} p_a,   \nonumber
\ee
we can represent this universal solution as follows,
\be
\gamma(x,p) = G(\hat{p}),  \quad\quad  G(z) := \frac{\ii \, z}{2} + \frac{z}{2} \cot{\frac{z}{2}}\,.\label{gammauniv}
\ee
Also the second master equation~\eqref{master2} exhibits the universal solution for $\rho$, expressed as a matrix valued function of
 the same matrix variable $\hat{p}$.
The explicit formula reads~\cite{Kupriyanov:2022ohu},
\be
\rho(x,p) = F(\hat{p}),\quad F(z)= \frac{e^{\ii z} -1}{\ii \, z}\,. \label{rhouniv}
\ee
These universal solutions for $\gamma$ and $\rho$ exhibit a simple connection~\cite{Kupriyanov:2022ohu},
\be
\rho^{-1} = \gamma - \ii\,\hat{p}.
\ee
Substituting the universal solutions~\eqref{gammauniv} and~\eqref{rhouniv}  in the formulae of Sec.~2, we can build the Poisson gauge model completely: 
we know the deformed gauge transformations~\eqref{infgtr}, which close the Poisson gauge algebra~\eqref{poga}, and we also know the gauge-covariant equations of motion~\eqref{nateqm}.
\subsection*{Example.}
\noindent In order to illustrate the formalism, we consider
the $\kappa$-Minkowski non-commutativity, which arises in various contexts, see e.g.~\cite{Mathieu:2020ywc,Lizzi:2021rlb,Mathieu:2021mxl,Meljanac:2022qed,Meljanac:2021qgq, HersentKilian:2023tnt} and for the recent progress.
The Poisson bivector, which corresponds to the (generalised) $\kappa$-Minkowski non-commutativity, is defined in the following way,
\be
\Theta^{ab} = 2(\omega^a x^b - \omega^b x^a)\,, \label{kappaPBV}
\ee
where $\omega^{a}$, $a=0,...,n-1$, stand for the deformation parameters.
The corresponding structure constants read,
\be
f^{ab}_c  = 2(\omega^a\delta^b_c - \omega^b\delta^a_c)\,.
\ee
The universal solutions have been calculated in~\cite{Kupriyanov:2022ohu},
\begin{eqnarray}
\gamma^{a}_b (A) &=&  (\omega\cdot A) \left[1+   \coth{(\omega\cdot A)} \right]\delta^a_b  +\frac{1-(\omega\cdot A)-(\omega\cdot A)\coth{(\omega\cdot A)}}{\omega\cdot A}\,\omega^a A_b\,,\nonumber\\
 \rho_b^a(A)   &=& \frac{e^{2(\omega\cdot A)}-1}{2(\omega\cdot A)}\, \,\delta^a_b + \frac{1+2(\omega\cdot A)-e^{2(\omega\cdot A)}}{2(\omega\cdot A)^2}\,\,\omega^a A_b\, . \label{kappanew}
 \end{eqnarray}
These expressions are different from the ones, obtained in the previous studies~\cite{Kupriyanov:2020axe}, using different techniques,
\begin{eqnarray}
\tilde\gamma^a_b(A)&=& \left[\sqrt{1+(\omega\cdot A)^2}+(\omega\cdot A)\right]\delta^a_b -\omega^a\,A_b \,,  \nonumber \\
\tilde\rho^a_b(A) &=&\left[\sqrt{1+(\omega\cdot A)^2}+(\omega\cdot A)\right]\,\delta_b^a -\frac{\sqrt{1+(\omega\cdot A)^2}+(\omega\cdot A)}{\sqrt{1+(\omega\cdot A)^2}}\,\omega^a A_b\, . \label{kappaold}
\end{eqnarray}
Therefore we have more than one Poisson gauge model, which corresponds to the Poisson bivector~\eqref{kappaPBV}.
In the next section we take a closer look at the arbitrariness of the construction.

\section{Arbitrariness and Seiberg-Witten maps.}
\noindent Consider a given Poisson bivector $\Theta^{ab}(x)$. Let $\gamma(x,p)$ and $\rho(x,p)$ be solutions of the master equations~\eqref{master1} and~\eqref{master2}, which have the commutative limits~\eqref{gcomlim} and~\eqref{rhocomlim} respectively. 
For any invertible field redefinition, 
\be
A\to\tilde A(A), \label{fredef}
\ee 
obeying the condition,
\be
\lim_{\Theta\to 0} \tilde A(A) =A, 
\ee
the quantities
 \be
\tilde{\gamma}^{a}_b (x,\tilde{A}) = \Bigg(\gamma_c^a(x,A)\cdot \frac{\partial \tilde{A}_b}{\partial A_c}\Bigg)\Bigg|_{A = A(\tilde{A})},\quad  \tilde{\rho}_a^i(x,\tilde{A})  = \Bigg(\frac{\partial A_s}{\partial \tilde{A}_i}\cdot \rho_a^s(x,A)\Bigg)\Bigg|_{A = A(\tilde{A})}\,,  \label{newoldgammarho}
 \ee
 are again the solutions of the master equations, which have correct commutative limits, see~\cite{Kupriyanov:2021aet,Kupriyanov:2022ohu} for details.  Therefore, the matrices $\tilde{\gamma}(x,p)$ and $\tilde{\rho}(x,p)$ define one more Poisson gauge model for the same Poisson bivector $\Theta^{ab}(x)$.
 
 The infinitesimal gauge transformation of the new model read, 
 \begin{equation}
\tilde\delta_f\tilde A_a(x)=\tilde\gamma^i_a(x,\tilde A(x))\,\partial_i f(x)+   {\{\tilde A_a(x),f(x)\}}\,. 
\end{equation}
Upon the field redefinition the gauge orbits are mapped onto the gauge orbits~\cite{Kupriyanov:2021aet,Kupriyanov:2022ohu}, 
\begin{equation}
\tilde A\left(A+\delta_ fA\right)=\tilde A(A)+\tilde\delta_f\tilde A(A)\,, 
\end{equation}
thus the invertible field redefinition~\eqref{fredef} defines a Seiberg-Witten map between the two models.
\subsection*{Example. } 
\noindent In the $\kappa$-Minkowski case, the universal solutions~\eqref{kappanew} and the  solutions~\eqref{kappaold}, which have been previously obtained in~\cite{Kupriyanov:2020axe}, are connected through the relation~\eqref{newoldgammarho} and the following Seiberg-Witten map~\cite{Kupriyanov:2022ohu},
\begin{equation}
\tilde A_a=\frac{\sinh (\omega\cdot A)}{\omega\cdot A} \,A_a\, \quad \Leftrightarrow \quad A_a=\frac{\mathrm{arcsinh}\,(\omega\cdot \tilde{A})}{\omega\cdot \tilde{A}} \,\tilde A_a\,.\nonumber 
\end{equation}

\section{Summary and concluding remarks.} 
\noindent 
The main points are the following.
\begin{itemize}
\item{For a given Poisson bivector $\Theta^{ab}$, defining the non-commutativity,  the main constituents of the Poisson gauge formalism, viz the deformed gauge transformations, the deformed field strength and the deformed gauge-covariant derivative, are completely determined by the matrices $\gamma$ and $\rho$, which solve the two master equations.}
\item{The Poisson gauge theory exhibits an elegant symplectic-geometric description in terms of symplectic embeddings and constrains in the extended space.}
\item{There exist universal solutions of the master equations, which are valid for \emph{all} non-commutativities of the Lie algebraic type. }
\item{Invertible field redefinitions give rise to new solutions of the master equations. All the outcoming Poisson gauge models are  connected with each other through Seiberg-Witten maps. }
\end{itemize}
Of course, our mini-review does not cover all aspects of Poisson gauge theory. In particular, an important connection with the 
$L_{\infty}$-algebras has not been discussed. The interested reader is referred to~\cite{Kupriyanov:2021cws} and~\cite{Abla:2022wfz}.
See also~\cite{Jonke:2022wpx,Szabo:2022edp} for applications of the $L_{\infty}$-structures to generalised gauge symmetries and non-commutative gravity.

In conclusion, we would like to outline a few open questions. First, one has to generalise the present results, introducing the charged matter.
Second, it would be very interesting to go beyond the semi-classical approximation towards the full non-commutative gauge algebra~\eqref{ncgalg}. Third, one may wonder what are the space-time symmetries of Poisson gauge models. In particular, one may study the fate of the discrete symmetries. In addition to the obvious phenomenological interest\footnote{See for instance~\cite{Novikov:2022pff} and references therein.} in the discrete symmetries breaking, the preserved symmetries such as PT allow the non standard quantum field theories that may play the important role in the dark energy physics~\cite{Andrianov:2016ffj,Novikov:2017dsl}.
Some aspects of the $PT$-symmetry in the non-commutative geometric context have already been studied in~\cite{Novikov:2019kit}. 

\subsubsection*{Acknowledgments} 
\noindent
The author is grateful to Vlad Kupriyanov and Patrizia Vitale for discussions and collaboration on related papers.

\end{document}